\documentstyle[twocolumn,aps]{revtex}
\begin{document}
\draft
\title{Electronic phase transitions in one-dimensional spinless fermion model with
competing interactions}
\author{A. K. Zhuravlev and M. I. Katsnelson\cite{E}}
\address{Institute of Metal Physics, Ekaterinburg 620219, Russia}
\author{A. V. Trefilov}
\address{Kurchatov Institute, Moscow 123182, Russia}
\maketitle

\begin{abstract}
An accurate numerical consideration is carried out of the ground state for
the simplified model which is traditionally used for the description of
Verwey transition and related phenomena. In the framework of 1D spinless
fermion model, the effects of next-nearest neighbour (NNN) interaction on
the metal-insulator transition are investigated for electron concentrations
1/2 and 2/3. It is shown that for large enough NNN transfer integrals the
electronic topological transition of metal-metal type is also possible. The
corresponding phase diagrams are presented.
\end{abstract}




\section{Introduction}

Studies of highly correlated electron systems are one of the ''hot'' areas
of condensed matter physics. A tremendous number of papers have been
concerned with the problems such as heavy fermions, Mott transitions,
magnetism of highly correlated systems, unconventional mechanisms of
superconductivity etc. In this context the classical problem of charge
ordering which has been raised by Wigner \cite{ref1} and Shubin and
Vonsovsky \cite{ref2} (see also \cite{ref3}) received far less attention. As
early as the 1930s, the famous Verwey transition in magnetite Fe$_3$O$_4$ at
the temperature $T_V=120K$ has been discovered. According to the traditional
views, this is the transition from the low-temperature insulating phase with
Fe$^{3+}$ and Fe$^{2+}$ ions forming superlattice (Wigner crystal) to the
high-temperature conducting phase with the destroyed charge ordering \cite
{ref4}. However, recent experimental data point to a more complicated
physical picture since the long-range charge ordering below $T_V$ appeared
to be pronounced rather weakly \cite{ref5,ref6} and the infrared spectra
show ''tunneling modes'' which are presumably connected with the
''smearing'' of Fe$^{3+}$ and Fe$^{2+}$ ions over finite clusters \cite{ref7}%
. The assertion that namely the charge ordering of Fe$^{3+}$ and Fe$^{2+}$
ions is the cause of the metal-insulator transition is also called in
question in a number of experimental works \cite{ref12,ref13}. Phenomena
similar to the Verwey transition are observed also in a number of other
compounds, e.g., in R$_3$X$_4$ (R=Sm,Eu; X=S,Se), Yb$_4$As$_3$ (see \cite
{ref8} and references therein), in the layered compounds RFe$_2$O$_4$ ,where
R is a rare-earth ion \cite{ref9}. Some of these compounds (e.g., Eu$_3$S$_4$%
) exhibit long-range charge ordering at low temperatures, and other ones
(e.g., Sm$_3$Se$_4$) show only short-range ordering. In the latter case, a
number of physical properties appeared to be anomalous, e.g., a giant
low-temperature heat capacity which is not connected with current carriers
is observed \cite{ref10,ref11}. Irkhin and Katsnelson \cite{ref8} suggested
that an RVB (or pseudospin liquid) -type state with separated pseudospin and
current degrees of freedom may be formed in such systems. Hence, the problem
of description of the Verwey transition and related phenomena seems to be
rather interesting from both experimental and theoretical points of view and
may turn out to be closely related to the more popular areas of the physics
of highly correlated electron systems.

\section{The formulation of the model}

To consider the effects of strong correlations in crystals including charge
ordering the polar model \cite{ref2} is usually used with the Hamiltonian

\begin{equation}  \label{eq1}
H=-\sum_{ij\sigma }{\rm ^{^{^{\prime}}}}t_{ij}c_{i\sigma
}^{\dagger}c_{j\sigma }+U\sum_in_{i\uparrow }n_{i\downarrow }+ {\frac 12}%
\sum_{ij}{\rm ^{^{^{\prime}}}}V_{ij}n_in_j {\ ,}
\end{equation}
where $c_{i\sigma }^{+}$ and $c_{i\sigma }$ are the creation and
annihilation electron operators, $i$ is the site (Wannier state) index, $%
\sigma =\uparrow ,\downarrow $ is the spin projection; $n_{i\sigma
}=c_{i\sigma }^{+}c_{i\sigma };n_i=n_{i\uparrow }+n_{i\downarrow }$ is the
operator of electron number on site $i$; $t_{ij}$ is the transfer integral; $%
U,V_{ij}$ are the matrix elements of Coulomb interaction on-site and between
different sites correspondingly, and the primed summation symbol indicates
that $i\neq j$. In comparison with the original polar model this expression
does not contain the multielectron contributions to the transfer processes
and the direct exchange interaction (see \cite{ref3} ). The effects of
electron-phonon interaction may be partially taken into account by the
renormalization of the model parameters (e.g. the formation of small-radius
bipolarons by the condition $V<0$ etc.).

In the case of Verwey transition in the magnetite the excess electrons
distinguishing a Fe$^{2+}$ site from a Fe$^{3+}$ one are almost completely
spin-polarized ($T_V<<T_C$ where $T_C=850K$ is the Curie temperature) and
fill the band by less than half. Therefore the term with $\sigma =\downarrow 
$ may be omitted in the Hamiltonian (1) and one may pass to the so-called
spinless fermion model

\begin{equation}
H=-\sum_{ij}{\rm ^{^{^{\prime}}}}t_{ij}c_i^\dagger c_j + {\frac{1}{2}}%
\sum_{ij}{\rm ^{^{^{\prime}}}}V_{ij}n_in_j.  \label{eq2}
\end{equation}

In the limit $U\rightarrow \infty $ the Hamiltonian (1) coincides with (2) 
\cite{ref25,ref26}. More exactly, this is valid only in 1D case since in 2D
and 3D systems the transfer Hamiltonian is renormalized by the spin-polaron
effects \cite{br}. But if we deal with ferromagnets (this is the magnetite
case) the latter are absent.

The present work is devoted to the rigorous numerical investigation of the
ground state of the model (2) in 1D case with the allowing for the
interaction and transfer processes for the next-nearest neighbors. As it
will be shown below they lead to a number of novel effects in particular to
the opportunity of dimerization or electronic phase transitions of
''metal-metal'' type. 1D model may be considered not only as a simplified
way to understand some features of real 3D case but also as a model to
describe the processes in the conducting polymers and other
quasi-one-dimensional crystals \cite{ref33,ref34,ref35}.

At $t_{ij}=0$ the model (2) is equivalent to the Ising model and describes
in the ground state a ''frozen'' electron distribution (Wigner crystal)
depending on the specific form of $V_{ij}$. At $V_{ij}=0$ we have the case
of noninteracting electrons with the homogeneous distribution. Intuitive
considerations of energy gain for competing processes show that at $%
t_{ij}\sim V_{ij}$ the ground state should change.

In the nearest-neighbor approximation 1D model with the Hamiltonian (2)
appears to be exactly soluble \cite{ref27,ref28}. For the electron
concentration per site $\rho =1/2$ and $V=2$ (here and further we put $%
t_{i,i\pm 1}=1,t_{i,i\pm 2}=t^{\prime },V_{i,i\pm 1}=V,V_{i,i\pm
2}=V^{\prime }$) the metal-insulator transition takes place with the
appearance of the energy gap in the electron spectrum at $V>2$. Let us
investigate for the beginning the effect of the interaction $V^{\prime }$ on
this transition at $t^{\prime }=0.$

For $t=0$ and $\rho =1/2$ the ground state in the model (2) is determined by
the ratio of $V$ and $V^{\prime }$ and, as it may be easily demonstrated,
corresponds to the ''usual'' charge ordering [...10101010...] for $V^{\prime
}<\frac 12V$ and dimer ordering [...110011001100...] for $V^{\prime }>\frac 1%
2V$ (1 means the electron and 0 means the hole on a site). So the model (2)
is frustrated because of the competing nature of the interactions $V$ and $%
V^{\prime }$ (they favor different types of charge ordering).

Let us define the quantity $\Delta _\infty $ as follows: it is equal to 0 if
the ground state degeneracy $g$ is macroscopically large (i.e. increase
faster than $N$ when $N\rightarrow \infty )$ at $t=0$; otherwise it is equal
to the difference between the energies of the lowest excited state and the
ground one at $t=0$. It may be shown by the direct enumeration of possible
states with the same energy that at $t=0$ the degeneracy $g$ is
exponentially large in $N$ at $\rho =1/2,V^{\prime }=\frac 12V$ and
therefore $\Delta _\infty =0$ in this case.

One may obtain for $\rho =1/2$

\begin{equation}
\Delta_\infty = \left\{ 
\begin{array}{ll}
{V - 2V^{\prime}} & \mbox{if ${1\over2}V>V'$,} \\ 
&  \\ 
{2V^{\prime}- V} & \mbox{if ${1\over2}V<V'<V$,} \\ 
&  \\ 
{V^{\prime}} & \mbox{if $V'>V$.}
\end{array}
\right.
\end{equation}

and for $\rho =2/3$

\begin{equation}
\Delta_\infty = \left\{ 
\begin{array}{ll}
{V^{\prime}} & \mbox{if $V>V'$,} \\ 
&  \\ 
{V} & \mbox{if $V<V'$.}
\end{array}
\right.
\end{equation}

As the electron transfer is taken into account the ground state for the case 
$\Delta _\infty =0$ is split into the energy band with the bandwidth of
order of $Nt.$ Since $g>>N$ this band should be continuous in the limit $%
N\rightarrow \infty $. Therefore one should expect that no energy gap will
appear between the ground state and the lowest excited one so the ground
state will be always conducting at $\Delta _\infty =0.$ At $\Delta _\infty
>0 $ and $t>0$ it is naturally to expect that the conducting ground state
will transform to the insulating one with the increasing $\Delta _\infty $.
The following question is to be elucidated: at what value of $\Delta _\infty
^c(V,V^{\prime })$ does the metal-insulator transition takes place and does
the transition point depend on $V$ and $V^{\prime }$ separately?

Hence, the following is the list of questions to be answered: (i) Does the
metal-insulator transition exist at $\Delta _\infty =0?$ (ii) At $\rho =1/2,$
does the metal-insulator transition occur at $\Delta _\infty \approx 2$
similarly to the exactly soluble case $V^{\prime }=0?$ (iii) More generally,
do the interaction constants enter the expression for the transition point
mainly via the parameter $\Delta _\infty ?$ (iv) How does the
metal-insulator transition proceed at $\rho \neq 1/2$ in particular at $\rho
=2/3$ (this filling corresponds to the situation in many real compounds, see
e.g. \cite{ref8})?

These problems were considered already in a number of papers (although, to
our knowledge, for the case $\rho =1/2$ only). In \cite{emery} the phase
diagram of the 1D spinless fermion model (rewritten in terms of spin
operators \cite{ref27}) with next-nearest-neighbor interaction has been
obtained by the renormalization group approach. In \cite{zotos,condmat} the
metal and insulator regions of this model has been investigated by an exact
numerical calculation of ''Drude'' contribution to the frequency dependent
conductivity (rather general approach to the numerical calculation of
correlation functions for 1D many-electron systems see e.g. also in \cite
{pang,poil}). However, the renormalization group method can give only
general shape of the phase diagram and not the exact numerical results for
the positions of phase boundaries. On the other hand, the calculations in 
\cite{zotos} have been carried out only for few sets of parameters of the
model. The recent work \cite{condmat} contains the most detailed numerical
investigation of the metal-insulator phase diagram in the model under
consideration. Below, the problem of metal-insulator transition in the model
(2) will be investigated by the calculations of the total energy and energy
gap by the Lanczos method for the broader range of the parameters than in
the previous papers and the answers to the above questions will be sought
for. Also the interesting phenomenon of the metal-metal transition which
appears to be possible in the model with next-nearest-neighbor transfer
processes will be studied.

\section{Metal-insulator transition: the results of calculations}

To investigate the ground state of the system we have used the Lanczos
method of exact diagonalization for finite clusters \cite{ref31,ref32}
(really we made the calculations for $N\leq 20$) with the extrapolation to
the limit $N\rightarrow \infty $. The comparison of the results with the
exact ones at $t^{\prime }=0,V^{\prime }=0$ \cite{ref27,ref28} have shown
that we have at least four accurate digits in the ground state energy and
two ones in the energy gap (which was calculated as the average difference
of the ground state energy at the adding and removing the electron). Such
accuracy is sufficient to answer the questions formulated above. The main
results are the following:

1. At $\Delta _\infty =0$ the metal-insulator transition is absent and the
ground state is always metallic both for $\rho =1/2$ and for $\rho =2/3$
(the calculations have been carried out up to $V\sim 10^2$).

2. The metal-insulator phase diagram for $\rho =1/2$ is shown in Fig.1. The
metal-insulator transition takes place at $\Delta _\infty \approx 2.$
Qualitatively this result seems rather natural. Nevertheless, it is not
trivial, to our opinion, that it turned out to be valid with the accuracy
comparable to the accuracy of the calculations. This result may be used for
the testing of different approximations proposed for the investigation of 2D
and 3D systems. Qualitatively our phase diagram is in agreement with that
from \cite{condmat}. Moreover, the part of the phase diagram for which real
calculations has been carried out in \cite{condmat} is the same. But our
results for the region of large $V^{\prime }$ are new. The main difference
is that, according to our results, the boundary lines of metal-insulator
transition are not crossed and the metallic phase is continued to the
infinity along the line $V^{\prime }=V/2$ in correspondence with the
qualitative discussion in the Introduction.

3. Fig.2 shows the phase diagram for $\rho =2/3$. The criterion of the
metal-insulator transition is $\Delta _\infty \approx 3.$ This result is new
even for the case $V^{\prime }=0$. The phase diagram for this case is
qualitatively different from that for $\rho =1/2$ (compare Figs.1 and 2). In
particular, the ground state is metallic for small enough $V^{\prime }$ and 
{\it arbitrary} $V$. It is connected with the macroscopically large
degeneracy of the ground state at $V^{\prime }=0.$ Also, it is not trivial
that the position of the metal-insulator boundary with the accuracy of the
calculations depend only on $V$ at $V^{\prime }>V$ and only on $V^{\prime }$
at $V>V^{\prime }$. Since the 1D spinless fermion model for $\rho =2/3$ was
probably not considered earlier we present in the Table 1 the results for
the ground state energy and the energy gap.

4. We also have carried out the calculations for the case $\rho =1/2$ with
the adding of the third-neighbor interaction term $V^{\prime \prime
}\sum_in_in_{i+3}.$ It appears that the criterion of the metal-insulator
transition $\Delta _\infty \approx 2$ is valid also in this case provided
that the interaction potential is downwards-convex namely $V+V^{\prime
\prime }>2V^{\prime },$ $\Delta _\infty $ being equal to

\begin{equation}  \label{eq12}
\Delta_\infty = \left\{ 
\begin{array}{ll}
{V - 2V^{\prime}+ 2V^{\prime\prime}} & \mbox{if ${1\over2}(V+V'')>V'$,} \\ 
&  \\ 
{2V^{\prime}- V} & \mbox{if ${1\over2}(V+V'')<V'<V$,} \\ 
&  \\ 
{V^{\prime}} & \mbox{if $V'>V$,}
\end{array}
\right.
\end{equation}

This statement is illustrated by the data of the Table 2: in the case of the
downwards-convex potentials (upper part of the Table) the criterion $\Delta
_\infty \approx 2$ holds but no simple criterion can be established for the
opposite case (lower part of the Table). Presumably, this affirmation is
also valid for the electron-electron interaction of arbitrary range. The
arguments justifying the downward-convexity condition for realistic
quasi-one-dimensional compounds were presented by Hubbard \cite{ref26}.

Apart from the metal insulator transition, another interesting phenomenon
studied in the spinless fermion model is the charge ordering. In particular,
for $\rho =1/2$ the interplay of ''usual'' and dimerized (Wigner and Peierls 
\cite{emery}) charge ordering can be discussed. Our approach cannot be
applied directly to the solution of this problem since it is rather
difficult to find asymptotics of the corresponding correlation functions
directly from finite-cluster calculations. This problem can be investigated
either by such approaches as analytical \cite{emery} or numerical \cite{pang}
renormalization group or by the combination of exact diagonalization
technique with the results of the theory of Luttinger liquid (see e.g. \cite
{poil}). Nevertheless we present here our results about the characteristics
of the {\it short-range} charge order in the model under consideration which
can be interesting themselves.

The results of the calculations for the correlation functions $\left\langle
n_0n_1\right\rangle $ and $\left\langle n_0n_2\right\rangle $ in the ground
state for $\rho =1/2$ are presented in Figs. 3 and 4. Consider first the
case $V^{\prime }<\frac 12V.$ According to Fig.3 in this case we always have 
$\left\langle n_0n_1\right\rangle <\left\langle n_0n_2\right\rangle $ which
show the absence of the dimerization. The metal-insulator transition is
almost not appreciable in the calculated correlation functions describing
the short-range order.

Consider now the case $V^{\prime }>\frac 12V$. The data presented in Fig. 4
show that as $V$ is decreased, the dimerized state with $\left\langle
n_0n_1\right\rangle >\left\langle n_0n_2\right\rangle $ is probably
destroyed before the metal-insulator transition occurs. One may suppose that
as the transfer integral $t$ is increased starting from the atomic limit,
first the dimer lattice melts and then the insulator-metal transition
occurs. Unfortunately the present results give us only preliminary
indications of the melting of dimerized lattice (because we cannot
investigate the asymptotics of the correlation functions and therefore have
no direct information about long-range order), and this interesting question
calls for further investigations.

The results of calculations for the correlation functions $\left\langle
n_0n_1\right\rangle ,$ $\left\langle n_0n_2\right\rangle $ and $\left\langle
n_0n_3\right\rangle $ for $\rho =2/3$ and $V=10$ is shown in Fig.5.
Qualitatively the similar picture takes place for any $V^{\prime }<V$. At
large $V$ and $V^{\prime }$ (or equivalently for $t\rightarrow 0$) we have $%
\left\langle n_0n_1\right\rangle \rightarrow 1/3,\left\langle
n_0n_2\right\rangle \rightarrow 1/3,\left\langle n_0n_3\right\rangle
\rightarrow 2/3.$

So our calculations give a rather full description of the metal-insulator
transition in the 1D spinless fermion model beyond the nearest neighbor
approximation. They also provide a basis for further investigations in
particular concerning the relation between the metal-insulator transition
and the destruction of the charge ordering. One may think that the derived
conclusions about the effects of competing interactions on the nature of
Verwey transition may be useful in analyzing the realistic 2D or 3D systems.

\section{Electronic topological transition in the model with
next-nearest-neighbor transfer processes}

Apart from the metal-insulator transition considered above the interelectron
interaction may lead to the electron phase transition of another type. It is
connected with the next-nearest neighbor transfer processes. Although the
role of such processes in many-electron models is investigated in a number
of works (see e.g. recent paper \cite{duffy}) the opportunity of such
transition, to our knowledge, was not noted and studied.

Consider at first the case $t^{\prime }\neq 0,V^{\prime }=0.$ In that case
at $V=0$ one-particle electron spectrum is given by

\begin{equation}  \label{spectrum}
\varepsilon_k = -2t \cos k - 2t^{\prime}\cos 2k .
\end{equation}

It is easily seen that at $t^{\prime }>t/4$ the spectrum is nonmonotonic in
the range $(0,\pi )$. At $t^{\prime }>t/2$ and $\rho =1/2$ the second
fermion ''pocket'' appears i.e. the electron-filled region in $k$ space is
no longer singly connected. In other words, while the Fermi surface of
''normal'' one-dimensional system consists of two points $k=\pm k_F$ it
consists of four points at $t^{\prime }>t/2$ and $\rho =1/2$. In the latter
case the occupied states corresponds to the intervals $-\pi \leq k\leq
-k_F^P,$ $-k_F^Q\leq k\leq k_F^Q,$ and $k_F^P\leq k\leq \pi .$ It should be
emphasized that 1D gas of non-interacting electrons cannot exhibit a
dispersion low of this type since it is well known (see e.g. \cite{ref36})
that the one-dimensional one-electron Schroedinger equation for a periodic
potential must necessarily exhibit a monotone $E(k)$ spectrum between $k=0$
and $k=\pi $ (the lattice parameter is set equal to unity everywhere).
Nevertheless, the Hamiltonian (1) with $t^{\prime }>t/4$ may be used for the
description of real systems e.g. consisting of the pairs of strongly coupled
one-dimensional chains with weak coupling between different pairs. In this
case $t$ is the transfer integral between the nearest sites in the direction
across the double chain and $t^{\prime }$ is the transfer integral along it
(see Fig. 6).

Studying of such systems is not only of purely theoretical interest. It may
give a deeper insight into the properties of realistic quasi-one-dimensional
systems in particular the well-known compound NMP-TCNQ (the molecules of
TCNQ and NMP are shown in Figs. 7a and 7b, respectively). This compound is
characterized by a charge transfer from TCNQ to NMP molecules (about 1/3
electron per NMP molecule). As one NMP-TCNQ cell contains one electron, half
of the $k$ states between $-\pi $ and $\pi $ are occupied. Therefore the
Fermi surface consists of two sheets: one of them is bounded by $k_F^Q=\pi
/3 $ and corresponds to the TCNQ chain; the other is bounded by $k_F^P=5\pi
/6$ and corresponds to NMP chain (see Fig. 7c) \cite{ref33}.

It is rather natural to discuss in such a case the following question: What
would happen to the distribution function upon switching the
electron-electron interaction, i.e., would the increasing interaction result
only in a gradual smearing of the steps in the distribution function of
electrons in $k$ space or a drastic change of its shape is also possible,
e.g., the merging of two steps? It will be shown below that the second
possibility really takes place.

Before the presentation and discussion of exact numerical results it would
be reasonable to treat this problem in the simplest Hartree-Fock
approximation. The trial wave function will be chosen in the form
corresponding to the distribution function with two steps in $k$ space of
the widths $2\pi \rho -2\lambda $ and $2\lambda $:

\begin{equation}  \label{eq4}
\vert \psi \rangle = \prod_k c_k^+ \vert 0 \rangle
\end{equation}
where quasimomentum takes on the values $-\pi \rho +\lambda \leq k\leq \pi
\rho -\lambda $ (the first step) and $-\pi \leq k\leq -\pi +\lambda ,\pi
-\lambda \leq k\leq \pi $ (the second step).

Substituting (7) into (2) and replacing the summation by the integration
according to the formula $\left( 1/N\right) \sum_k...\rightarrow \left(
1/2\pi \right) \int_{-\pi }^\pi dk...$ one has

\begin{eqnarray}
& \displaystyle \frac{E}{N} = \frac{\langle\psi\vert H\vert\psi\rangle}{N} =
\rho^2 V - \frac{2}{\pi} t [\sin (\rho\pi - \lambda) - \sin\lambda] - & 
\nonumber \\
& \displaystyle - \frac{1}{\pi} t^{\prime}[\sin (2\rho\pi - 2\lambda) + \sin
2\lambda] - \frac{V}{\pi^2} [\sin (\rho\pi - \lambda) - \sin\lambda]^2. &
\label{eq5}
\end{eqnarray}

In the case of $\rho =1/2,$ the substitution $x=1-\cos \lambda +\sin \lambda 
$ yields

\begin{equation}
\frac{E}{N} = \frac{V}{4} - \frac{2}{\pi} t + \frac{2}{\pi} ( t -
2t^{\prime}+ \frac{V}{\pi} ) x + \frac{1}{\pi} ( 2t^{\prime}- \frac{V}{\pi}
) x^2,  \label{eq6}
\end{equation}
where $0\leq x\leq 2,x=0$ corresponds to one pocket and $x>0$ corresponds to
two pockets. Minimizing the energy with respect to $x$ one obtains that the
increase of $V$ results in the merging of the two steps (disappearing of the
second pocket) at a critical value

\begin{equation}
V_c = \pi ( 2t^{\prime}- t ).  \label{eq7}
\end{equation}

This means that the switching on the electron-electron interaction,
according to the usual Hartree-Fock approximation, may lead to a topological
transition i.e. a change in topology of Fermi surface: the two-sheeted (four
point) ''surface'' transform into a one-sheeted (two-point) one. In a 3D
case it would correspond to a transition from a doubly connected Fermi
surface to a singly connected one which is a particular case of the
electronic topological transitions proposed by Lifshitz \cite{ref37}. The
investigation of such opportunity by a more rigorous way is the purpose of
our calculations described below.

The calculations have been carried out for a system of 10 electrons in the
ring from 20 sites and of 8 electrons in the ring from 16 sites. The results
for the points of the electronic topological transition in these two sets of
calculations were the same with the accuracy of two significant digits. The
computational results are shown in Figs. 8,9. Fig.8 displays the
distribution function in the ground state $\left\langle n_k\right\rangle $
at $t^{\prime }=t$. It is clearly seen that ''the hump'' near the edges of
the Brillouin zone ($k=-\pi ,\pi $) disappears at $2.5t<V<2.6t$. Fig.9
presents the phase diagram of the system. The comparison with Eq.(10) show
that beginning from approximately $t^{\prime }\approx 0.8t$ the Hartree-Fock
approximation gives too high values for $V_c$ and the difference with
numerical results grows with increasing $t^{\prime }$. So, the Hartree-Fock
approximation is not very accurate quantitatively at large $V$.

In the case $V^{\prime }\neq 0$ the inverse topological transition turns out
to be possible when the increase of the Coulomb interaction results in the
appearance of the second pocket. We have carried out the corresponding
calculations for the case $t^{\prime }=0.49t,.\rho =1/2.$ Thus, without
interaction the second minimum in the electron spectrum lies very close to
the Fermi level and above it. The interaction can make this minimum to be
lower than the Fermi level. The phase diagram for this case is shown in
Fig.10. The Hartree-Fock approximation can describe qualiatively the
phenomenon of ''inverse'' transition but gives too high values of the
critical values of $V^{\prime }$ (the difference may be in a factor of order
of 2-5 depending on $V$ values).

To conclude this section note why these results are, to our opinion,
non-trivial. It is well-known that for 1D systems the electron velocity may
be zero either at $k=0$ or at the boundary of the Brillouin zone \cite{ref36}%
. Therefore, true one-dimensional systems cannot exhibit Van Hove
singularities inside the allowed band and hence they exhibit no electronic
topological transitions because the latter are nothing but the crossing of
the Fermi level by the Van Hove singularity. However, this is not the case
for, e.g., a double chain. The exact numerical results presented here
suggest that in this case quasi-one-dimensional systems do may demonstrate
electronic topological transitions. As in the 3D case \cite{ref37} they are,
unlike metal-insulator transitions, transitions of metal-metal type.
Studying the possibility of such transitions in real quasi-one-dimensional
conductors containing double chains would be of experimental interest.

We are grateful to A.O.Anokhin and S.V.Tretjakov for the assistance in
performing of numerical calculations.

CAPTIONS\ TO\ FIGURES

Fig.1. Metal-insulator phase diagram at $\rho =1/2$. Black and empty circles
correspond to the metal and insulating phase, respectively; solid lines
correspond to the condition $\Delta _\infty =2.$

Fig.2. Metal-insulator phase diagram at $\rho =2/3$. Black and empty circles
correspond to the metal and insulating phase, respectively; solid lines
correspond to the condition $\Delta _\infty =3.$

Fig.3. Dependences of the correlation functions $\left\langle
n_0n_1\right\rangle $ (circles) and $\left\langle n_0n_2\right\rangle $
(squares) on the parameter $V$ for different $V^{\prime }\leq V/2$: (1) $%
V^{\prime }=0$ (2) $V^{\prime }=0.2V$ (3) $V^{\prime }=0.3V$ (4) $V^{\prime
}=0.5V$.

Fig.4. Dependence of the correlation functions $\left\langle
n_0n_1\right\rangle $ (circles) and $\left\langle n_0n_2\right\rangle $
(squares) on the parameter $V$ for $V^{\prime }=0.7V.$

Fig.5. Dependences of the correlation functions $\left\langle
n_0n_1\right\rangle $ (circles), $\left\langle n_0n_2\right\rangle $
(squares) and $\left\langle n_0n_3\right\rangle $ (triangles) on $V^{\prime
} $ at $V=10t,\rho =2/3.$

Fig.6. The double chain.

Fig.7. (a) TCNQ molecule (hydrogen atoms are not shown) (b) NMP molecule (c)
the sketch of the dispersion law of electrons in NMP-TCNQ according to \cite
{ref33}.

Fig.8. The distribution function $\left\langle n_k\right\rangle $ at $%
t^{\prime }=t$. The solid line corresponds to $V=0,$ the dash-dot line
corresponds to $V=2.5t,$ the dashed line corresponds to $V=2.6t.$

Fig.9. Phase diagram of the 1D spinless fermion model at $\rho
=1/2,V^{\prime }=0,$ $\alpha $ being the label for ''one-pocket'' region and 
$\beta $ for ''two-pocket'' one.

Fig.10. Phase diagram of the 1D spinless fermion model at $\rho
=1/2,t^{\prime }=0.49t,$ $\alpha $ being the label for ''one-pocket'' region
and $\beta $ for ''two-pocket'' one.


\mediumtext
\begin{table}[tbp]
\caption{Dependence of the ground state energy (the upper lines) and the gap 
$\Delta$ (the lower lines) on the different $V$ and $V^{\prime}$ at $\rho
=2/3$}
\begin{tabular}{|c|c|c|c|c|c|c|c|c|c|}
$V^{\prime}$ & 0 & 1 & 2 & 3 & 4 & 5 & 6 & 8 & 10 \\ \hline
$V$ &  &  &  &  &  &  &  &  &  \\ 
0 & -0.5472 & -0.1318 & 0.2662 & 0.6489 & 1.0188 & 1.3790 & 1.7324 & 2.4248
& 3.1072 \\ 
& 0.000 & -0.001 & -0.005 & 0.004 & -0.004 & -0.003 & 0.002 & 0.004 & -0.001
\\ 
1 & -0.1856 & 0.2347 & 0.6386 & 1.0276 & 1.4031 & 1.7675 & 2.1236 & 2.8180 & 
3.5016 \\ 
& 0.000 & 0.000 & -0.003 & 0.004 & 0.003 & 0.002 & 0.001 & 0.001 & -0.001 \\ 
2 & 0.1669 & 0.5903 & 0.9984 & 1.3922 & 1.7728 & 2.1417 & 2.5010 & 3.1997 & 
3.8839 \\ 
& -0.001 & 0.000 & 0.000 & 0.002 & -0.004 & 0.002 & 0.003 & 0.003 & 0.004 \\ 
3 & 0.5135 & 0.9395 & 1.3495 & 1.7459 & 2.1292 & 2.5008 & 2.8626 & 3.5662 & 
4.2545 \\ 
& -0.002 & 0.000 & 0.000 & -0.001 & -0.004 & 0.005 & 0.005 & 0.005 & 0.006
\\ 
4 & 0.8565 & 1.2832 & 1.6941 & 2.0916 & 2.4757 & 2.8467 & 3.2072 & 3.9079 & 
4.5944 \\ 
& -0.002 & 0.000 & -0.003 & 0.005 & 0.094 & 0.140 & 0.169 & 0.092 & 0.083 \\ 
5 & 1.1971 & 1.6236 & 2.0355 & 2.4332 & 2.8166 & 3.1864 & 3.5454 & 4.2440 & 
4.9294 \\ 
& -0.002 & 0.000 & -0.001 & 0.020 & 0.309 & 0.610 & 0.876 & 1.018 & 0.932 \\ 
6 & 1.5367 & 1.9631 & 2.3746 & 2.7721 & 3.1547 & 3.5234 & 3.8814 & 4.5788 & 
5.2637 \\ 
& -0.003 & 0.000 & -0.003 & 0.019 & 0.374 & 1.017 & 1.497 & 1.952 & 1.934 \\ 
7 & 1.8744 & 2.3005 & 2.7123 & 3.1095 & 3.4913 & 3.8591 & 4.2164 & 4.9131 & 
5.5976 \\ 
& -0.003 & 0.000 & -0.004 & 0.039 & 0.422 & 1.316 & 2.042 & 2.803 & 2.925 \\ 
8 & 2.2113 & 2.6376 & 3.0488 & 3.4458 & 3.8271 & 4.1941 & 4.5510 & 5.2471 & 
5.9314 \\ 
& -0.004 & 0.000 & -0.001 & 0.066 & 0.586 & 1.470 & 2.447 & 3.443 & 3.822 \\ 
10 & 2.8830 & 3.3092 & 3.7205 & 4.1166 & 4.4969 & 4.8632 & 5.2192 & 5.9147 & 
6.5986 \\ 
& -0.004 & 0.000 & 0.000 & 0.082 & 0.570 & 1.390 & 2.943 & 4.568 & 5.423 \\ 
20 & 6.2307 & 6.6539 & 7.0639 & 7.4582 & 7.8362 & 8.2005 & 8.5555 & 9.2496 & 
9.9330 \\ 
& -0.004 & -0.003 & 0.000 & 0.127 & 0.748 & 1.793 & 3.036 & 5.781 & 8.721 \\ 
100 & 32.9074 & 33.3307 & 33.7389 & 34.1312 & 34.5074 & 34.8702 & 35.2243 & 
35.9175 & 36.6004 \\ 
& -0.005 & -0.003 & 0.015 & 0.173 & 0.861 & 1.963 & 3.292 & 5.983 & 8.813 \\ 
&  &  &  &  &  &  &  &  & 
\end{tabular}
\end{table}

\narrowtext
\begin{table}[tbp]
\caption{Dependence of the gap $\Delta$ on the parameter $\Delta_\infty$ at
different $V$, $V^{\prime}$ and $V^{\prime\prime}$ at $\rho =1/2$}
\begin{tabular}{|c|c|c|c|c|}
$V$ & $V^{\prime}$ & $V^{\prime\prime}$ & $\Delta_\infty$ & $\Delta$ \\ 
\hline
0.5 & 0 & 0.25 & 1.0 & 0.000 \\ 
2.0 & 1 & 0.50 & 1.0 & -0.001 \\ 
3.0 & 2 & 1.00 & 1.0 & -0.001 \\ 
4.5 & 3 & 1.50 & 1.5 & 0.038 \\ 
1.0 & 0 & 0.50 & 2.0 & 0.003 \\ 
4.0 & 2 & 1.00 & 2.0 & 0.096 \\ 
6.0 & 4 & 2.00 & 2.0 & 0.180 \\ 
7.5 & 5 & 2.50 & 2.5 & 0.465 \\ 
1.5 & 0 & 0.75 & 3.0 & 0.263 \\ 
6.0 & 3 & 1.50 & 3.0 & 0.726 \\ 
9.0 & 6 & 3.00 & 3.0 & 0.873 \\ 
2.0 & 0 & 1.00 & 4.0 & 1.069 \\ 
8.0 & 4 & 2.00 & 4.0 & 1.827 \\ 
10.0 & 5 & 2.50 & 5.0 & 3.119 \\ 
3.0 & 0 & 1.50 & 6.0 & 3.335 \\ 
4.0 & 0 & 2.00 & 8.0 & 5.296 \\ 
2.0 & 2 & 1.00 & 2.0 & 0.002 \\ 
3.0 & 3 & 1.50 & 3.0 & 0.000 \\ 
4.0 & 4 & 2.00 & 4.0 & 0.156 \\ 
5.0 & 5 & 2.50 & 5.0 & 1.769 \\ 
6.0 & 6 & 3.00 & 6.0 & 3.494 \\ 
&  &  &  & 
\end{tabular}
\end{table}


\begin{references}
\bibitem[*]{E}  E-Mail: mikhail.katsnelson@usu.ru

\bibitem{ref1}  E.Wigner, Trans. Farad. Soc. {\bf 34}, 679 (1938)

\bibitem{ref2}  S.P.Schubin and S.W.Wonsowsky, Phys. Zs. UdSSR {\bf 7}, 292
(1935); {\bf 10}, 348 (1936)

\bibitem{ref3}  S.V.Vonsovsky and M.I.Katsnelson, J.Phys.C{\bf 12}, 2042
(1979)

\bibitem{ref4}  N.F.Mott, {\it Metal-Insulator Transitions} (London, Taylor
and Francis, 1974)

\bibitem{ref5}  S.Iida, Phil. Mag. B{\bf 42}, 349 (1980)

\bibitem{ref6}  J.R.Cullen, Phil. Mag. B{\bf 42}, 387 (1980)

\bibitem{ref7}  P.Wachter and L.Degiorgi, Sol. State Comm.{\bf \ 66}, 211
(1988)

\bibitem{ref12}  K.P.Belov, Uspekhi Fiz. Nauk {\bf 163}, 53 (1993);
Zh.Eksp.Teor.Fiz. {\bf 110}, 2093 (1996)

\bibitem{ref13}  A.N.Goraga et al, Zh.Eksp.Teor.Fiz. {\bf 99}, 238 (1993)

\bibitem{ref8}  V.Yu.Irkhin and M.I.Katsnelson, Pis'ma ZhETF {\bf 49}, 500
(1989); Phys.\ Lett. A{\bf 150}, 47 (1990)

\bibitem{ref9}  M.Tanaka et al, J. Phys. Soc. Japan {\bf 58}, 1433 (1989)

\bibitem{ref10}  T.Furuno et al, J. Magn. Magn. Mater. {\bf 76/77}, 117
(1988)

\bibitem{ref11}  O.Nakamura et al, J. Magn. Magn. Mater. {\bf 76/77}, 293
(1988)

\bibitem{ref25}  A.A.Ovchinnikov, Zh.Eksp.Teor.Fiz. {\bf 64}, 342 (1973)

\bibitem{ref26}  J.Hubbard, Phys. Rev. B{\bf 17}, 494 (1978)

\bibitem{br}  W.F.Brinkman and T.M.Rice, Phys. Rev. B{\bf 2}, 1324 (1970)

\bibitem{ref33}  D.Jerome and H.J.Schulz, Adv. Phys. {\bf 31}, 299 (1982)

\bibitem{ref34}  A.J.Heeger, S.Kivelson, J.R.Schrieffer and W.-P.Su, Rev.
Mod. Phys. {\bf 60}, 781 (1988)

\bibitem{ref35}  J.Tsukamoto, Adv. Phys.{\bf \ 41}, 509 (1992)

\bibitem{ref27}  J. Des Cloizeaux and M.Gaudin, J. Math. Phys. {\bf 7}, 1384
(1966)

\bibitem{ref28}  J. Des Cloizeaux, J. Math. Phys. {\bf 7}, 2136 (1966)

\bibitem{emery}  V.J.Emery and C.Noguera, Phys. Rev. Lett. {\bf 60}, 631,
(1988)

\bibitem{zotos}  X.Zotos and P.Prelovsek, Phys. Rev. B{\bf 53}, 983 (1996)

\bibitem{condmat}  D.Poilblanc, S.Yunoki, S.Maekawa, and E.Dagotto,
cond-mat/9704170

\bibitem{pang}  H.Pang, H.Akhlaghpour, and M.Jarrell, Phys. Rev. B{\bf 53},
5086 ( 1996)

\bibitem{poil}  C.A.Hayward, D.Poilblanc and D.J.Scalapino, Phys. Rev. B{\bf %
53}, R8863\ (1996); C.A.Hayward and D.Poilblanc, Phys. Rev. B{\bf 53}, 11721
(1996)

\bibitem{ref31}  B.N.Parlett,{\it \ The Symmetric Eigenvalue Problem }%
(Englewood Cliffs, NJ, Prentice-Hall, 1980)

\bibitem{ref32}  E.Dagotto, Rev. Mod. Phys. {\bf 66}, 763 (1991)

\bibitem{duffy}  D.Duffy and A.Moreo, Phys. Rev. B{\bf 55}, R676 (1997)

\bibitem{ref36}  S.V.Vonsovsky and M.I.Katsnelson, {\it Quantum Solid Sate
Physics} (Berlin etc, Springer, 1989)

\bibitem{ref37}  I.M.Lifshitz, Zh.Eksp.Teor.Fiz. {\bf 38}, 1569 (1960)
\end{references}
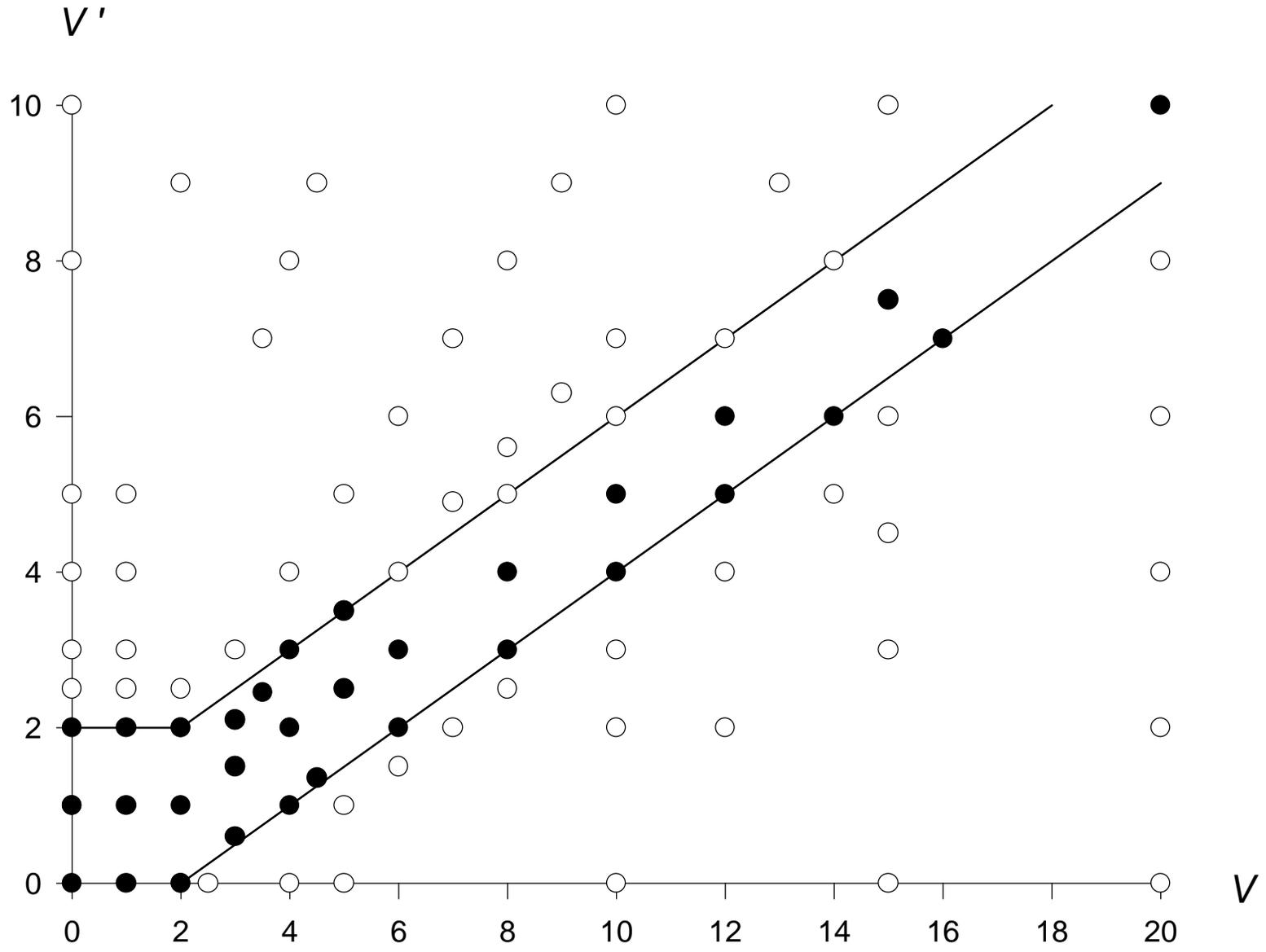
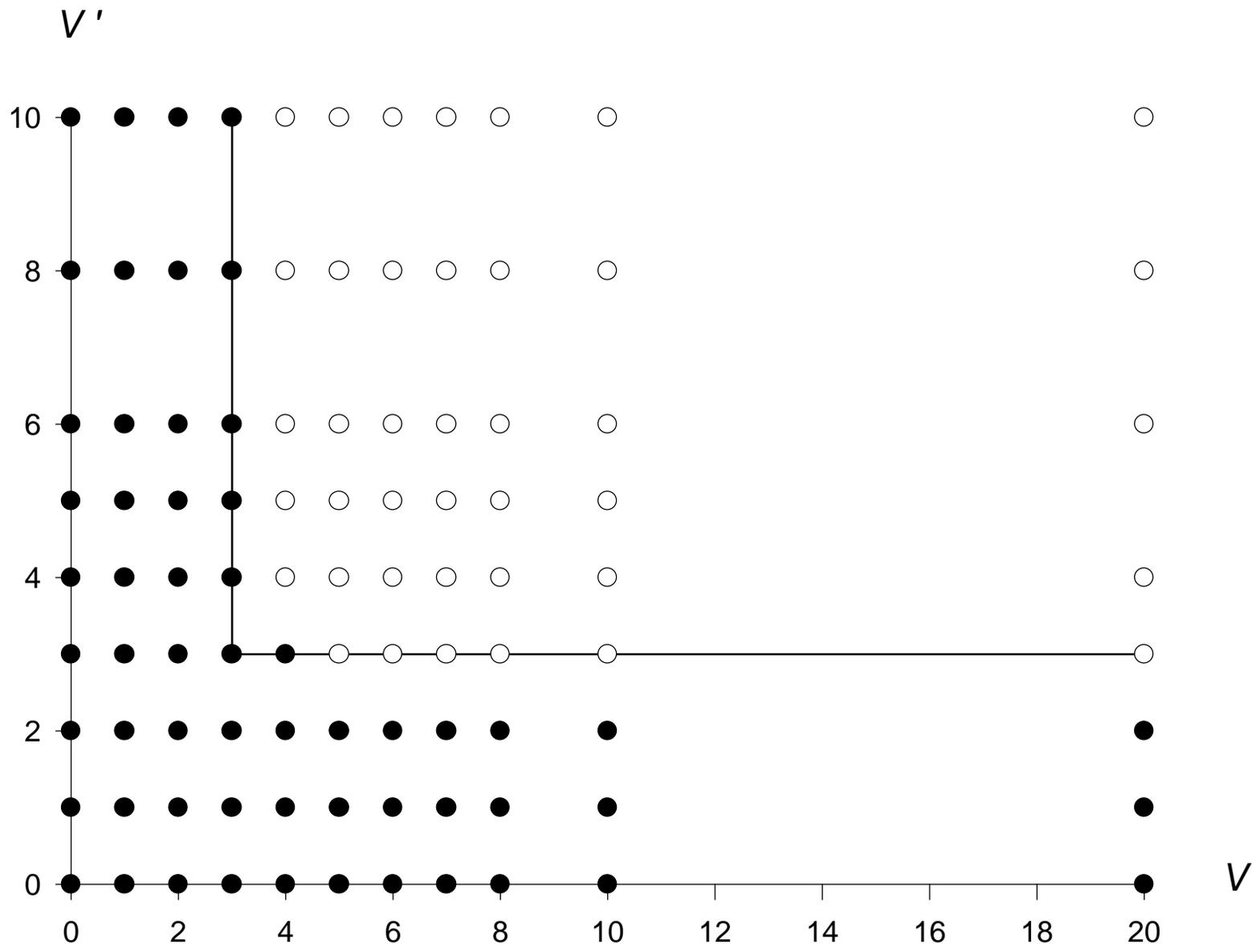
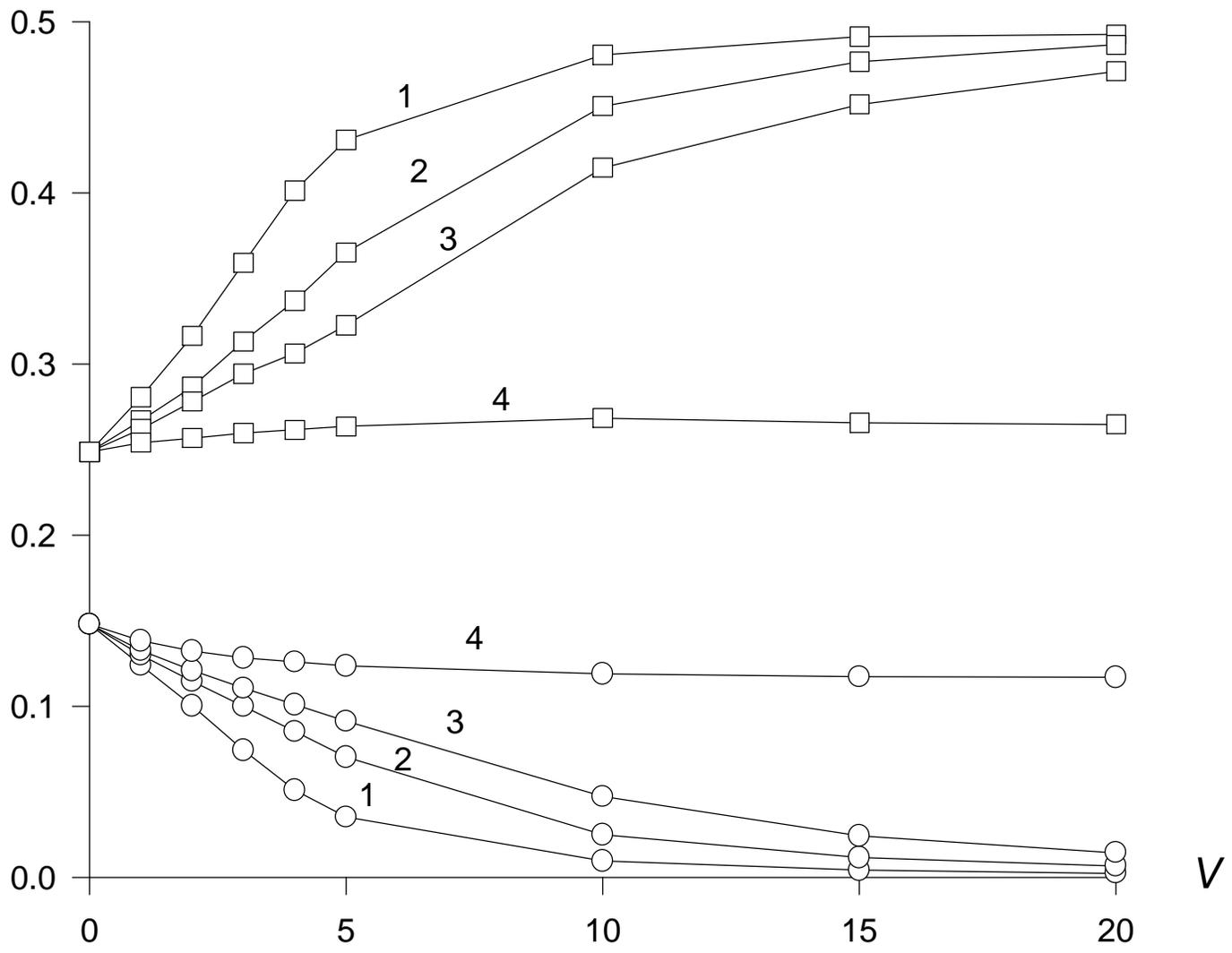
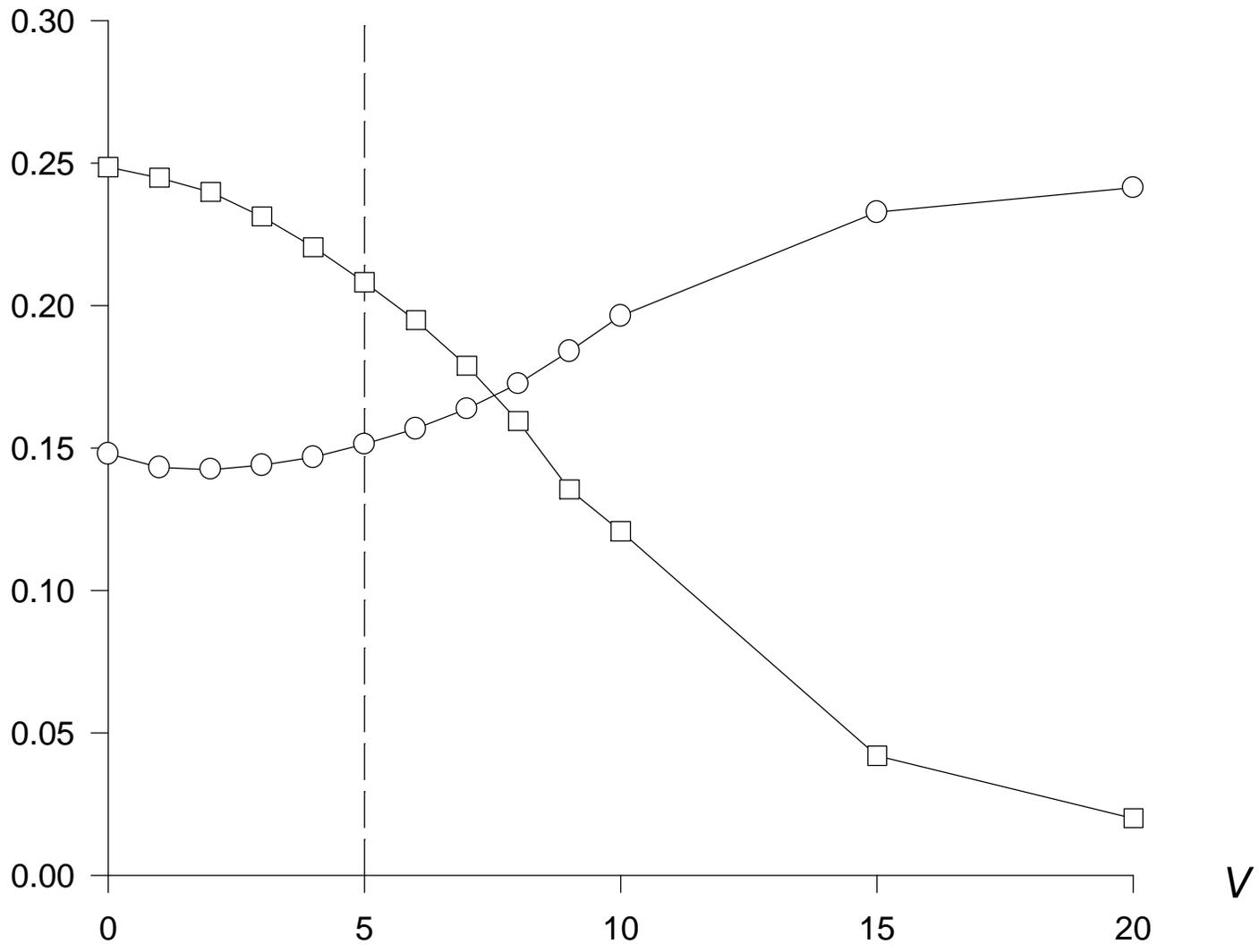
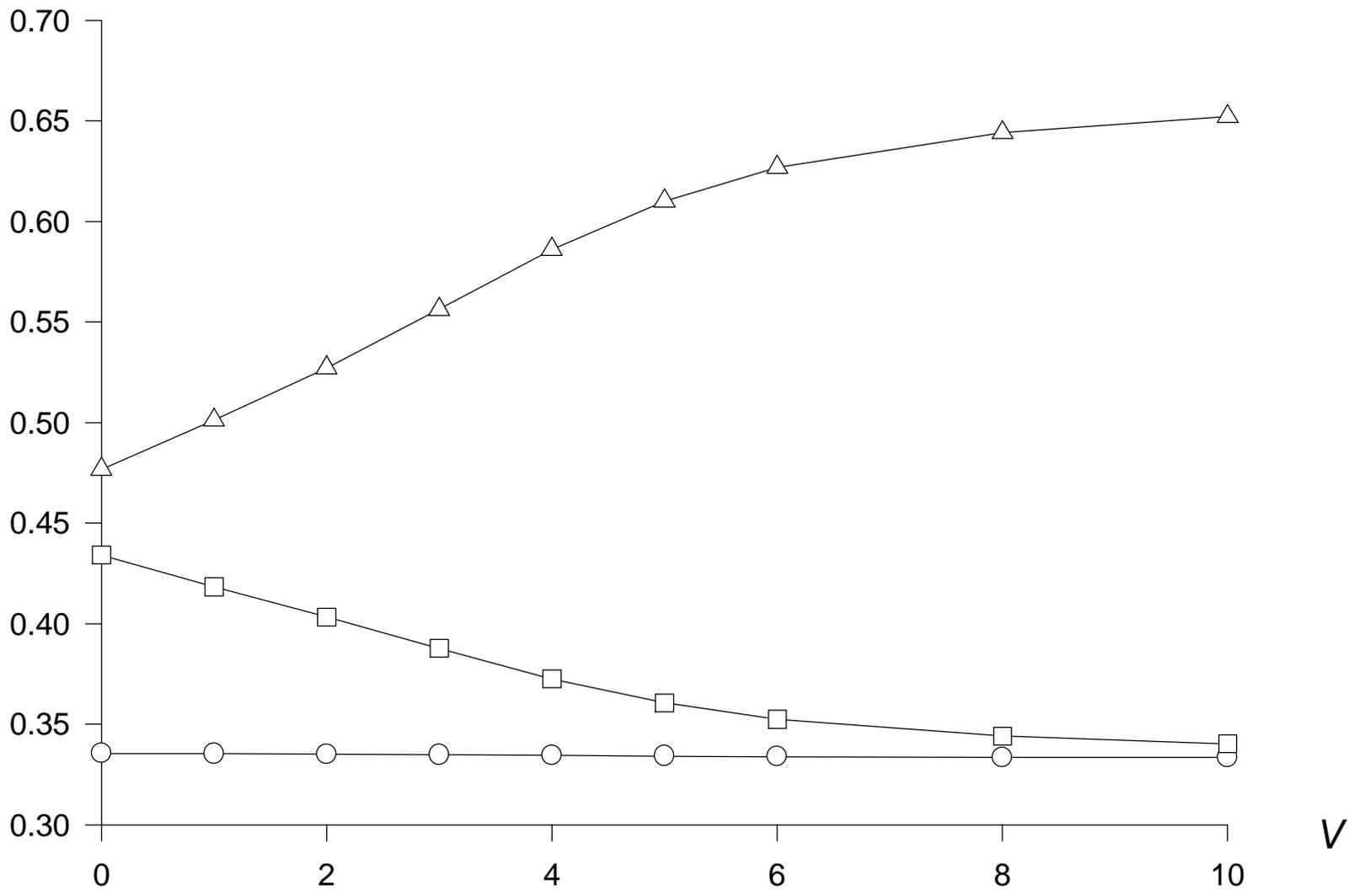
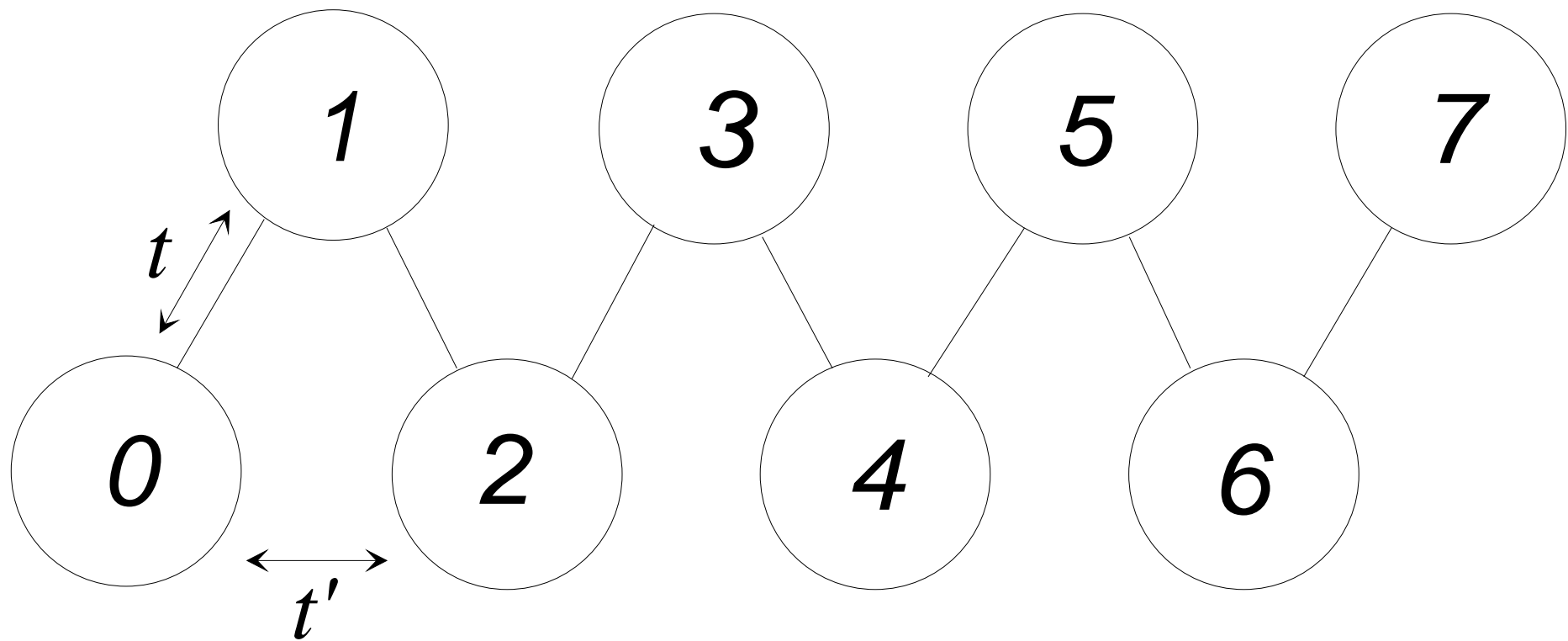
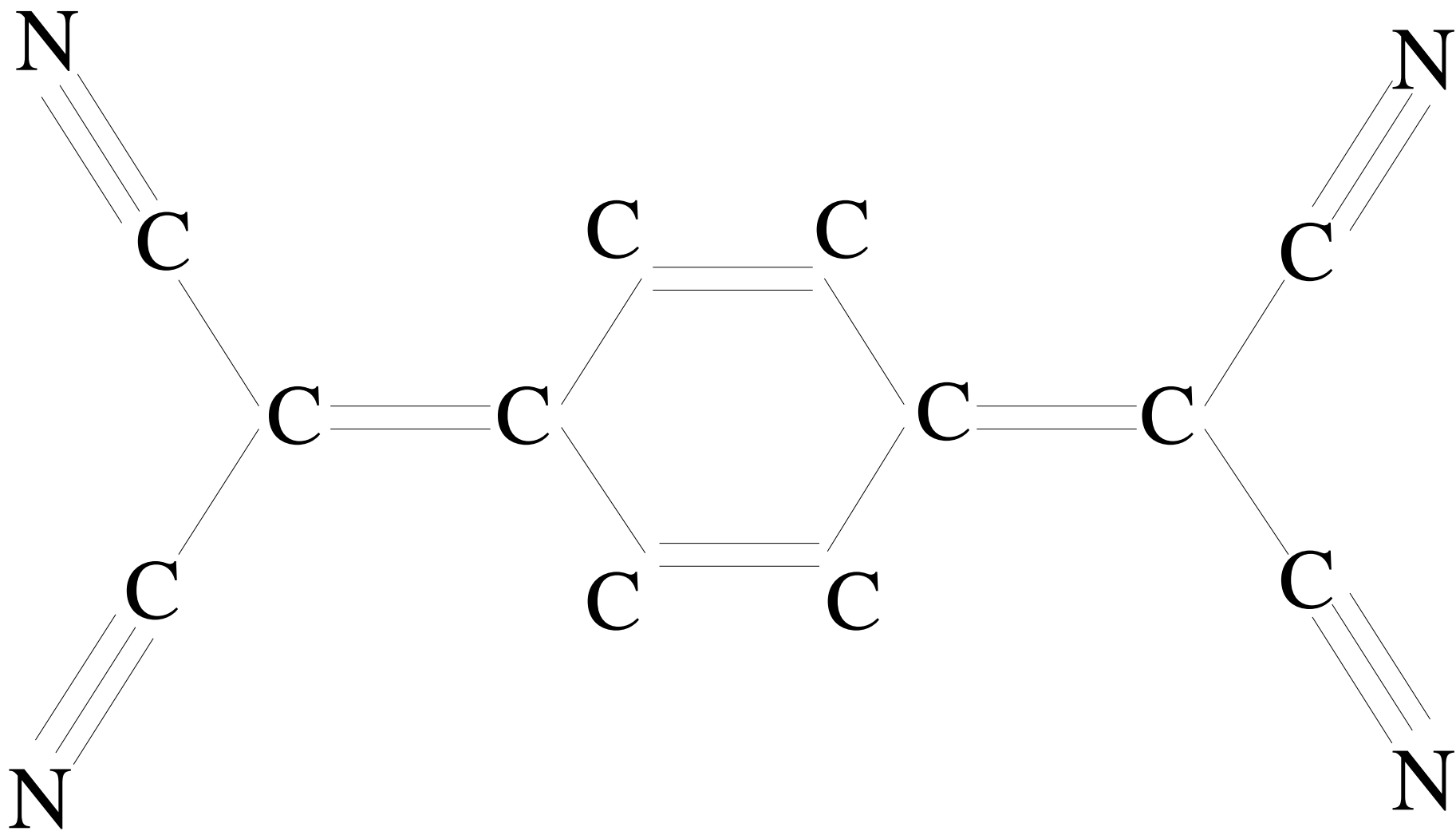
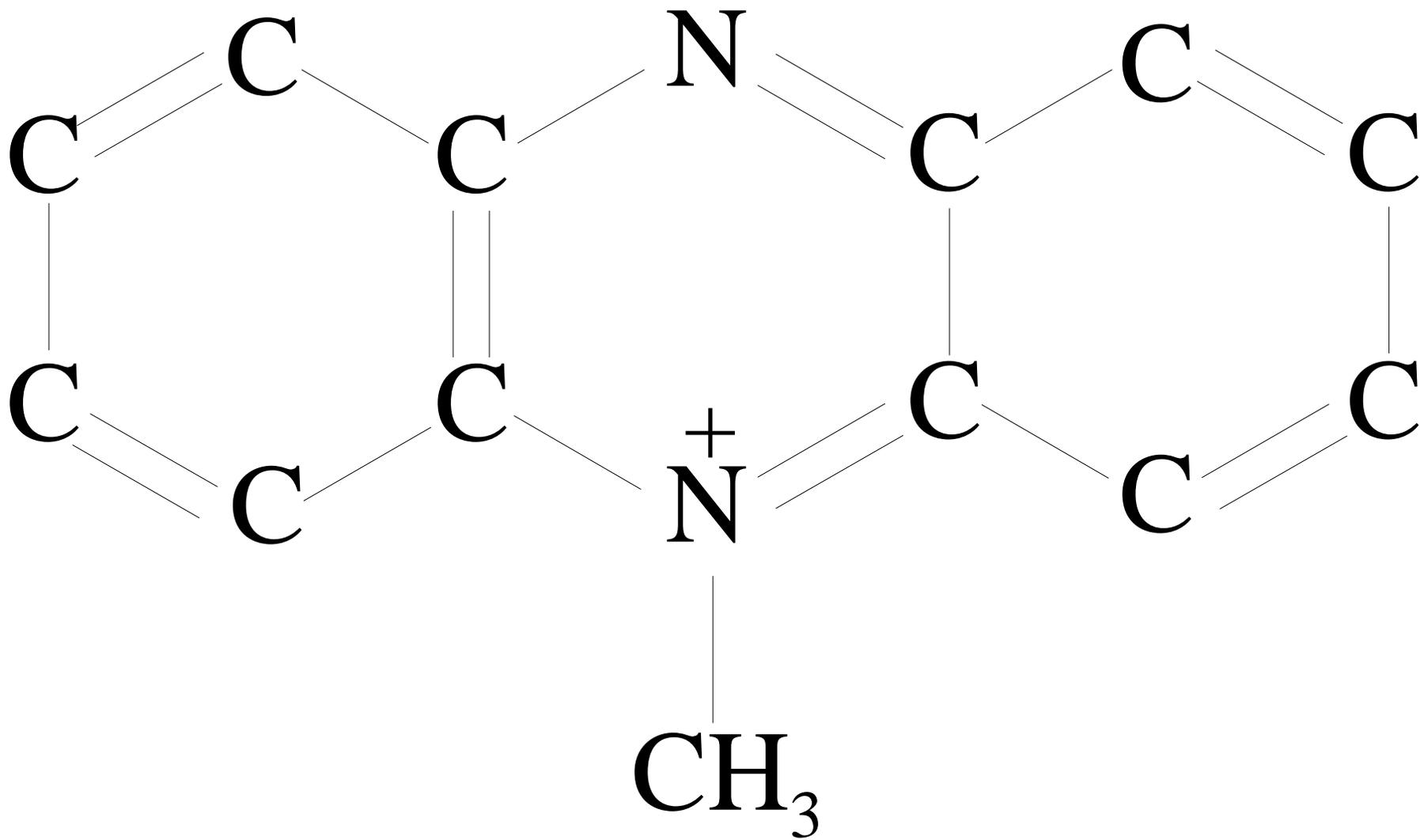
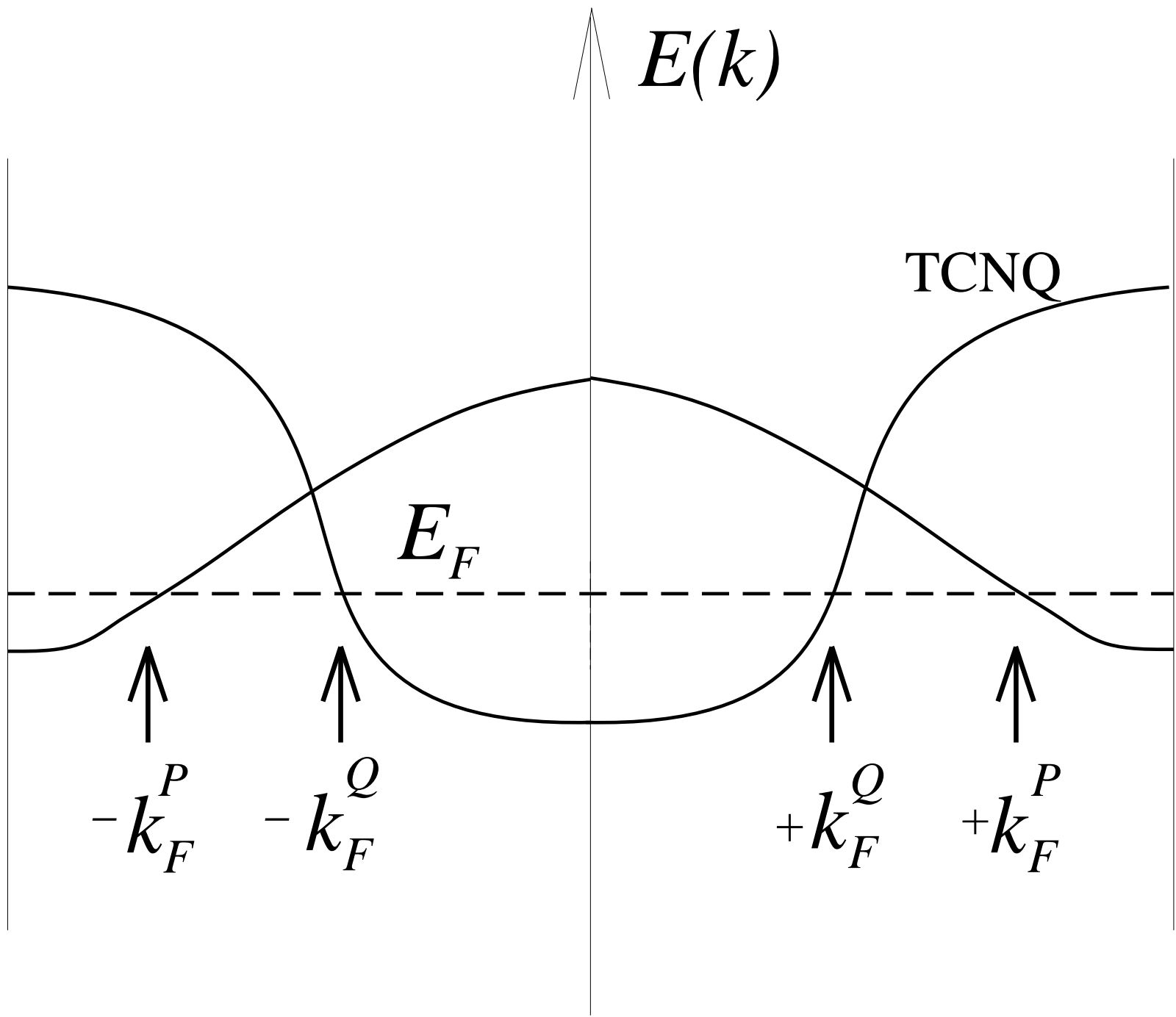
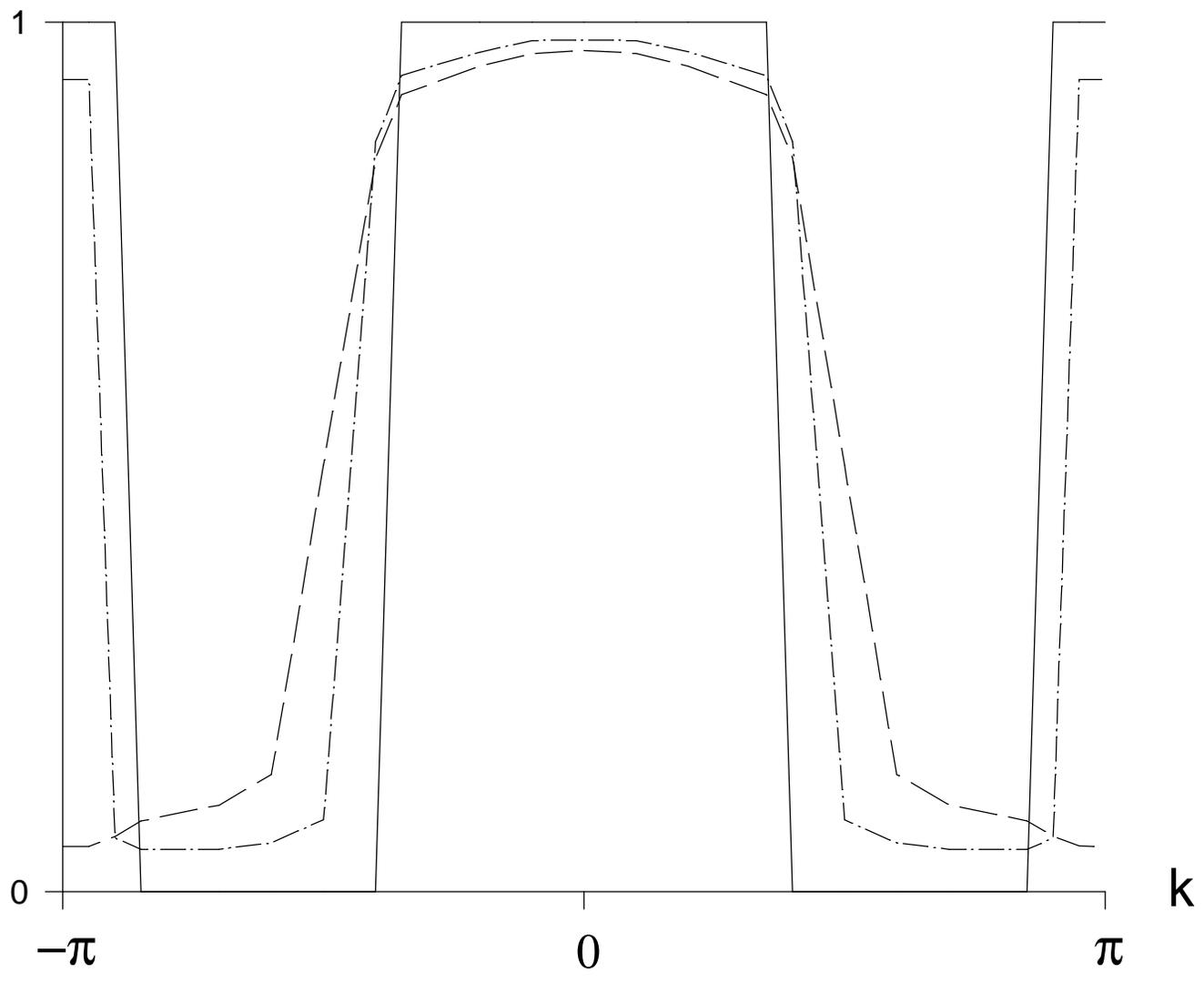
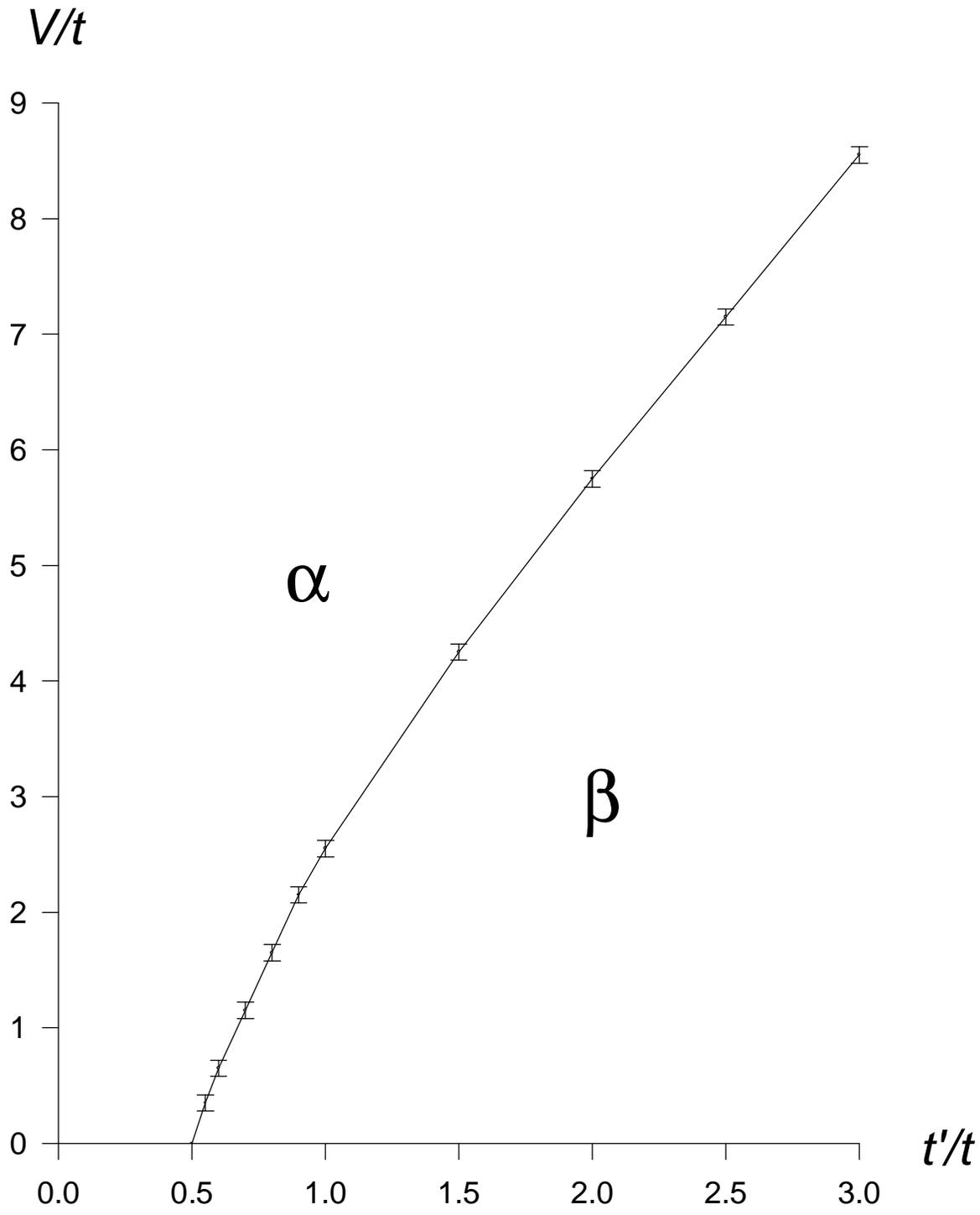
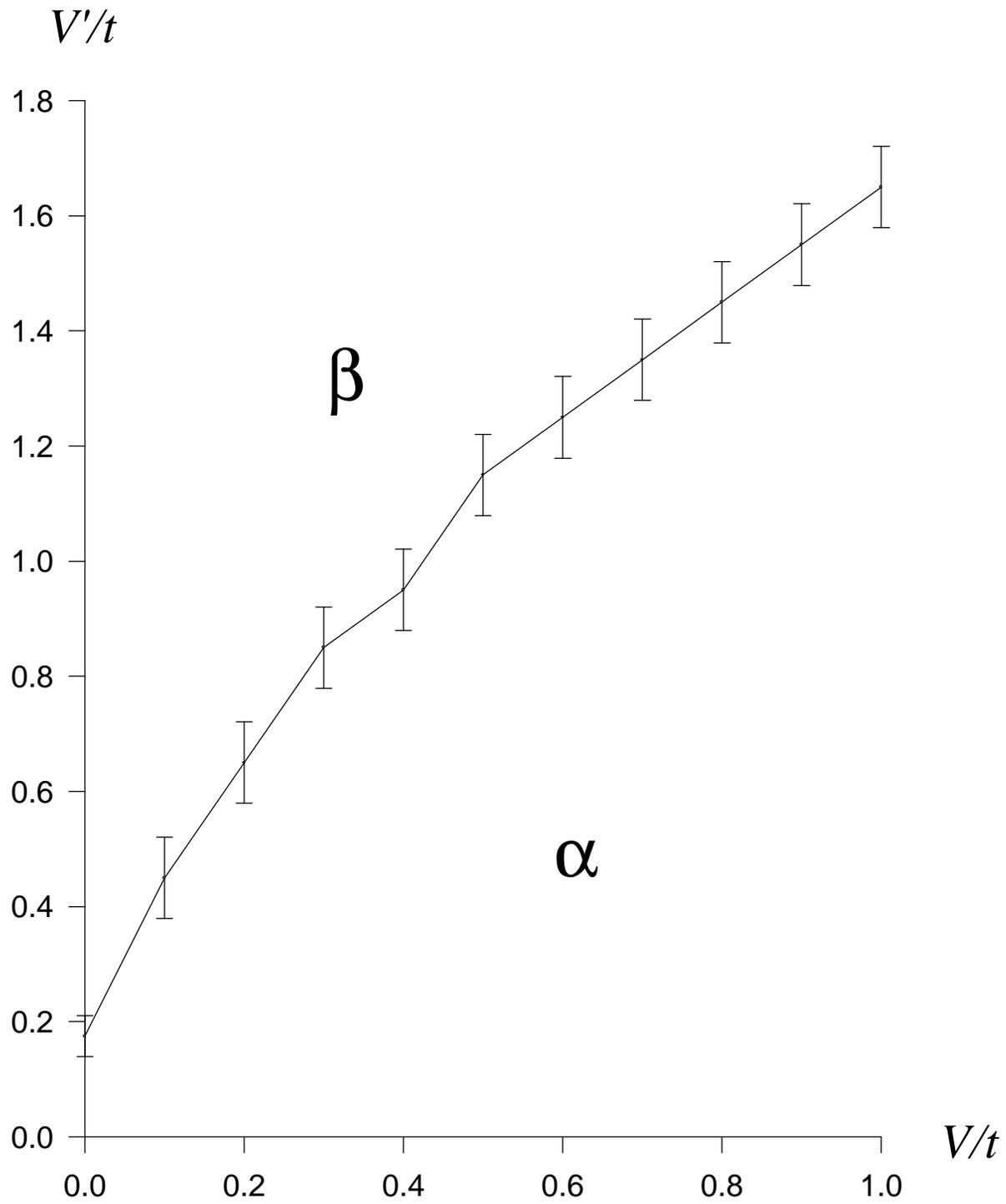
\end{document}